\documentclass[aps,prl,twocolumn,groupedaddress]{revtex4-1}

\begin{document}

\title{Strong-Field S-Matrix Theory With Coulomb-Volkov Final State in All Orders
}

\author{F.H.M. Faisal}
\email[]{ffaisal@physik.uni-bielefeld.de}
\affiliation{Fakult\"at f\"ur Physik, Universit\"at Bielefeld, D-33501 Bielefeld, Germany\\
\&\\
University of Arizona, Optical Sciences Center, Tucson, AZ 85721, USA}


\date{\today}

\begin{abstract}
Despite its long standing usefulness for the analysis of various processes in intense laser fields, it is well-known that the so-called strong-field KFR or SFA ansatz does not account for the final-state Coulomb interaction. Due to its importance for the ubiquitous ionisation process, numerous heuristic attempts have been made during the last several decades to account for the final state Coulomb interaction with in the SFA. Also to this end an ad hoc model with the so-called Coulomb-Volkov final state was introduced a long time ago. However, till now, no systematic strong-field S-matrix expansion using the Coulomb-Volkov final state could be found. Here we solve this long standing problem by determining the Coulomb-Volkov Hamiltonian, identifying the rest-interaction in the final state, and explicitly constructng the Coulomb-Volkov propagator (or Green's function). We employ them to derive the  complete S-matrix series for the ionisation amplitude governed by the Coulomb-Volkov final state in all orders. The results are given in both ``velocity'' and ``length'' gauges. We also present a gauge independent version of the Coulomb-Volkov S-matrix series.
\end{abstract}

\pacs{32.80.Rm,42.50.Hz, 34.80.Qb
}

\maketitle

\section{Introduction}
Over the past several decades the well-known strong-field approximation in the form of the so-called KFR or SFA  ansatz \cite{kel,faia,rei}
has provided much fruitful insights into the highly non-perturbative processes that occur during light-matter interaction in intense laser fields.
However, it is also well-known that SFA, based as it is on the plane-wave Volkov state, does not account for the Coulomb interaction in the final state. The latter, however, is specially significant for the ubiquitous ionisation process. Due to this difficulty, many authors in the past decades have made various {{\it ad hoc}} corrections to SFA. One such attempt was to introduce a long time ago \cite{jai,fer}
the so-called Coulomb-Volkov state and a heuristic one-term ionisation amplitude using it. However, till now, no systematic strong-field S-matrix expansion accounting for the final state Coulomb interaction in terms of the Coulomb-Volkov wavefunction  
has been found. Here we solve this long standing problem and derive a systematic 
S-matrix series with the Coulomb-Volkov final-state in all orders. To this end, we first determine: (i) the exact Coulomb-Volkov Hamiltonian, (ii) the complete set of linearly independent fundamental solutions of the Coulomb-Volkov Schroedinger equation and, (iii) the associated Coulomb-Volkov propagator (or Green's function). They are employed systematically to derive the S-matrix series of the ionisation amplitude. The result is presented in both the so-called ``velocity" and the ``length" gauges. We end the paper by presenting  also a gauge independent version of the Coulomb-Volkov S-matrix series.

\section{Three-interaction Formalism}
For the present purpose we shall use below a three-interaction formalism developed earlier in connection with the so-called intense-field 
S-matrix theory or IMST (see, e.g. review \cite{bec_fai} or, original references cited therein). First, we outline the three-interaction technique suitable for the problem at hand.

The Schroedinger equation of the interacting atom+ laser field is 
 \begin{equation}\label{TotSchEqu}
(i\hbar \frac{\partial}{\partial t} - H(t))| \Psi(t)\rangle=0
\end{equation}
where $H(t)$ is the total Hamlltonian of the system,
\begin{equation}
H(t)= H_a + V_i(t)
\end{equation}
For example, for an effective one electron atomic system interacting with a laser field, we may take 
\begin{eqnarray}\label{AtoHam}
H_a &=& (\frac{{\vec{p}_{op}}^2}{2m} - \frac{Ze^2}{r}+ V_{s.r.}(\vec{r}) )\nonumber\\
\end{eqnarray}
where $Z$ is the core charge and $V_{s.r.}(\vec{r})$ is a short-range potential that goes to zero for asymptotically large $r$ faster than the Coulomb potential.

The laser-atom interaction is assumed here in the minimal coupling gauge (in ``dipole" approximation) 
\begin{eqnarray}\label{IniIntVG}
V_i(t)&=& (-\frac{e}{mc}\vec{A}(t) \cdot\vec{p}_{op}+ \frac{e^2A^2(t)}{2mc^2})\nonumber\\
\end{eqnarray}
where $\vec{A}(t)$ is the vector potential of the laser field, and $\vec{p}_{op}\equiv -i\hbar\nabla$.

Since all information of the interacting system is contained in the full wavefunction $\Psi(t)$ and in general this is not known explicitly,
we shall consider a more useful formal expression of 
the full wavefunction in terms of the appropriate partial interactions among the sub-systems and, the associated sub-propagators
(or Green's functions). The latter objects may be already known or could be found to expand the total wavefunction.

Thus, first, we may formally define the full propagator, $G(t,t')$, associated with the total Hamiltonian  
$H(t)$, by the inhomogeneous equation
\begin{equation}\label{TotGreEqu}
(i\hbar \frac{\partial}{\partial t} - H(t)) G(t,t')= \delta(t-t').
\end{equation}
The solution of the Schr\"odinger equation (\ref{TotSchEqu}) 
can then be expressed as 
\begin{eqnarray}\label{FulKet}
|\Psi(t)\rangle &=& |\phi_i(t)\rangle +  \int G(t,t_1) V_i(t_1)|\phi_i(t_1)\rangle dt_1
\end{eqnarray}
where $|\phi_i(t)\rangle$ is a given initial state. 
We may note here already that due to the implicit presence of the Heaviside theta-function in all the propagators (see, for example, the Volkov propagator given in the sequel) the time integration limits are always from a given initial time $t_i$ to a given final time $t_f$.  The limits of the intermediate time-integrations are automatically controlled by the propagators at the appropriate positions by themselves. Usually the full interaction time interval, $t_f- t_i$, is taken to be long, e.g., from $-\infty$ to $+\infty$. Note, however, that there is no difficulty in using the theory for a finite or a very short interaction time e.g. with ultra short laser pulses, for during the rest of the time (from the lower to the upper long-time limits) the pulse could be assumed to be vanishingly small.

In general, as for the full wavefunction, we do not have explicit knowledge of the full propagator $G(t,t')$; therefore, we intend to re-express it in terms of certain relevant sub-propagators that are already known or would be found.
Clearly, the two most relevant states in any quantum mechanical transition process are the initial state in which the system is prepared and the final state in which the system is detected. Since in any ionisation process the final state interaction is governed by the long-range Coulomb interaction of the outgoing electron and the residual ion-core, it is highly desirable that the final state incorporates the long-range Coulomb interaction from the beginning.
Let us define a final reference Hamiltonian $H_f(t)$ that incorporates the final-state Coulomb interaction in the presence of the laser field.
Formally, the final state propagator is defined as usual by 
\begin{equation}
( i\hbar \frac{\partial}{\partial t} - H_f (t))G_f(t,t')(t) =\delta(t-t')
\end{equation}
Assuming for a moment that a suitable $H_f(t)$ and $G_f(t,t')$ for the present purpose could be found, 
the total $G(t,t')$ can then be re-expressed, in terms of $G_f(t,t')$, as
\begin{eqnarray}\label{GinGf}
G(t,t')&=& G_f(t,t') + \int G_f(t,t_1)V_f(t_1)G(t_1,t') dt_1
\end{eqnarray}
Substituting this in $|\Psi(t)\rangle$ above we get a ``closed" form expression of the full wavefunction in the form
\begin{eqnarray}\label{FulWavFun}
|\Psi(t)\rangle &=& |\phi_i(t)\rangle +  \int dt_1G_f(t,t_1) V_i(t_1)|\phi_i(t_1)\rangle\nonumber\\
&+&\int dt_2 dt_1 G_f(t,t_2) V_f(t_2) G(t_2,t_1) V_i(t_1)|\phi_i(t_1)\rangle\nonumber\\
\end{eqnarray}
This form of the wavefunction (or state vector) of the interacting system has been originally derived and discussed in connection with  double ionisation processes (see, e.g. review [4]). Here we shall make use of it for the problem at hand.
In fact, the transition amplitude (or the S-Matrix element $S_{fi}$) from an initial state, $|\phi_i(t)\rangle$, to a final state $\langle\psi_f(t)|$ of the system is given, by definition, by the projection of the final state on to the total wavefunction evolving from the initial state. Thus, using the above form of $|\Psi(t)\rangle$, we get
\begin{eqnarray}\label{SMatSer0} 
S_{fi} & =& \langle \psi_f(t)|\Psi(t)\rangle\nonumber\\
&=& \langle \psi_f(t)|\phi_i(t)\rangle+ \int dt_1 \langle \psi_f(t_1)| V_i(t_1)|\phi_i(t_1)\rangle \nonumber\\
&+& \int dt_2 dt_1 \langle\psi_f(t_2) |V_f(t_2) G(t_2,t_1) V_i(t_1)|\phi_i(t_1) \rangle\nonumber\\
& +& ...
\end{eqnarray}
This is a specially convenient general form of a transition amplitude from which to generate the desired expansion of the ionisation amplitude. 
Now, $G(t,t')$ may be expanded in terms of {\it any} suitable intermediate sub-propagator and the corresponding intermediate interaction (without affecting the choice of the initial and the final states and the respective rest-interactions).  Here we choose the strong-field Volkov propagator $G_{Vol}(t,t')$ to expand the full $G$ appearing in the intermediate position in the expression above.

The Volkov Hamiltonian is given by the interaction of the free-electron with the laser field only, or
\begin{equation}
H_{Vol}(t)=({\frac{{\vec{p}}_{op}^2}{2m} -\frac{e}{mc}\vec{A}}(t)\cdot{\vec{p}} +\frac{e^2 A^2(t)}{2mc^2} )
\end{equation}
The solution of the corresponding Schoedinger equation is easily found 
\begin{equation}\label{VolSol}
\psi_{\vec{p}}(\vec{r},t)=\langle\vec{r}|\vec{p}\rangle e^{-\frac{i}{\hbar} \int^t p_{t'}^{2}/(2m) dt^{'}}
\end{equation}
where ${\vec{p}_t}\equiv(\vec{p}-\frac{e}{c}\vec{A}(t))$ and
$\langle \vec{r}|\vec{p}\rangle = e^{\frac{i}{\hbar}\vec{p}\cdot\vec{r}}$ is a plane wave of momentum $\vec{p}$.\\

The Volkov propagator $G_{Vol}(t,t')$ is the solution of the inhomogeneous equation
\begin{equation}
(i\hbar\frac{\partial}{\partial t} - ({\frac{{\vec{p}}_{op}^2}{2m} -\frac{e}{mc}\vec{A}}(t)\cdot{\vec{p}} +\frac{e^2 A^2}{2mc^2} )) G_{Vol}(t,t')= \delta(t-t')
\end{equation}
In terms of the Volkov states it is given explicitly by:     
\begin{equation}\label{VolPro}
G_{Vol}(t,t') = -\frac{i}{\hbar}\theta(t-t')\sum_{\vec{p}} \frac{1}{L^3}
|\vec{p}\rangle e^{-\frac{i}{\hbar} \int_{t'}^t \frac{p_{t"}^2}{2m}  dt"}
\langle\vec{p}|
\end{equation}
Using the Volkov propagator we can expand 
\begin{eqnarray}\label{VolExpG}
G(t,t')&=& G_{Vol}(t,t') + \int G_{Vol}(t,t_1)V_0(t_1)G_{Vol}(t_1,t') dt_1\nonumber\\
&+& \cdots.
\end{eqnarray}
The associated rest-interaction $V_0(t)$ is accordingly defined by 
\begin{eqnarray}\label{RestIntVol}
V_0(t)&=& H(t)-H_{Vol}(t)\nonumber\\
&=& (-\frac{Z e^2}{r} + V_{s.r}(\vec{r}) )
\end{eqnarray}
(which is time independent in the present case).

Since the initial state belongs to the atomic Hamiltonian $H_a$, therefore, the initial rest-interaction $V_i(t)$ is, as indicated earlier, simply
\begin{eqnarray}\label{IniInt}
V_i(t)&=&H(t)-H_a\nonumber\\
&=&(-\frac{e}{mc}\vec{A}(t)\cdot \vec{p}_{op}  +\frac{e^2 A^2}{2mc^2}   )
\end{eqnarray}

For the final state, we intend to take account of the long-range Coulomb interaction explicitly. One such state that takes  the final state Coulomb interaction into account is the so-called ``Coulomb-Volkov" state. It has been originally introduced a long time ago \cite{jai} by taking the usual stationary Coulomb-wave \cite{lan} and augmenting it heuristically by the time- dependent Volkov-phase:  
\begin{eqnarray}\label{CouVolSol}
\Phi_{\vec{p}}(\vec{r},t) &=& \phi_{\vec{p}}^{(-)}(\vec{r}) e^{-\frac{i}{\hbar} \int^t (\frac{p^2}{2m}-\frac{e}{c}\vec{A}(t)\cdot\vec{p}_{op} +\frac{e^2A^2(t)}{2mc^2}) dt' }
\end{eqnarray}
The stationary Coulomb waves, $\phi_{\vec{p}}^{(-)}(\vec{r})$, belong to the hydrogenic or, asymptotic Coulomb Hamiltonian $H_{Cou}$, 
\begin{eqnarray}
H_{Cou} &=& H_a - V_{s.r.}(\vec{r}) \nonumber\\
&=& (\frac{\vec{p}_{op}^2}{2m} -\frac{Ze^2}{r})
\end{eqnarray}
They are given by \cite{lan}
\begin{eqnarray}\label{CouWav}
\phi_{\vec{p}}^{(-)}(\vec{r}) &=& \frac{1}{L^{\frac{3}{2}}}e^{\frac{\pi}{2} \eta_p}\Gamma(1+ i \eta_p) 
e^{\frac{i}{\hbar} \vec{p}\cdot\vec{r}}\nonumber\\
&\times & { }_{1}F_{1}(- i \eta_p,1,-i (p r + \vec{p}\cdot\vec{r}))
\end{eqnarray}
We have assumed them to be normalised in a large volume $L^3$ with the understanding that,
limit $L \rightarrow \infty, \sum_{\vec{s}} (\cdots)\equiv (\frac{L}{2\pi})^{3} \int d^3s (\cdots)$;  
$\vec{p}_{op}\equiv -i\hbar \vec{\nabla}$, and
$\eta_p \equiv \frac{Z \hbar}{a_0 p}$ is the so-called Sommerfeld parameter;
$a_0=$ Bohr radius $=\frac{\hbar^2}{m e^2}$. Note that the ingoing ``minus" Coulomb wave is chosen above, which is appropriate for the ionisation final state. (The outgoing ``plus" wave, relevant  e.g. for the laser assisted scattering problem, is related to the ``minus" wave by
$\phi_{\vec{p}}^{(+)}(\vec{r})=\phi_{-\vec{p}}^{(-)*}(\vec{r})$.)

Note that the ansatz (\ref{CouVolSol}) does {\it not} fully satisfy the Schroedinger equation of the interacting system (\ref{TotSchEqu}). It is interesting, therefore, to ask: what is the Hamiltonian or the Schroedinger equation of which the ``Coulomb-Volkov" state, Eq. (\ref{CouVolSol}), is an exact solution? Essentially it is the lack of this information that has so far hindered the development of a systematic Coulomb-Volkov S-matrix theory where the final-state Coulomb interaction is taken account of through the Coulomb-Volkov state in all orders. Therefore, to proceed further we shall first determine the Coulomb-Volkov Hamiltonian (to be denoted $H_{CV}$ below) and the complete set of linearly independent solutions of the associated Schroedinger equation. This would allow us to construct both the Coulomb-Volkov propagator, $G_{CV}$, and to identify the rest-interaction in the final state with respect to the Coulomb-Volkov Hamiltonian $H_{CV}$.
 
\section{Coulomb--Volkov Hamiltonian and Propagator}

To determine the Hamiltonian $H_{CV}(t)$ to which the Coulomb-Volkov state belongs,
we introduce a vector operator
defined by
\begin{eqnarray}\label{PIOpe}
\vec{\pi}_{c} \equiv \sum_{\vec{s}} |\phi_{\vec{s}}\rangle\vec{s}\langle\phi_{\vec{s}}|
\end{eqnarray}
where 
$|\phi_{\vec{s}}\rangle$ stands for the Coulomb {\it continuum} waves with momentum $\vec{s}$ (cf. Eq. (\ref{CouWav})). 

Consider next the exponential operator
\begin{eqnarray}\label{TOpe1}
T(\vec{\pi}_{c}) = e^{i \vec{\alpha}(t)\cdot\vec{\pi}_c}
\end{eqnarray}
where $\vec{\alpha}(t) = \frac{e}{m c}\int^t\vec{A}(t')dt'$.
By expanding the exponential as a power series and using the projection operator nature of the individual terms, it can be reduced to
the simple form
\begin{eqnarray}\label{TOpe2}
T(\vec{\pi}_{c}) = 1 -\sum_{\vec{s}} |\phi_{\vec{s}}\rangle (1- e^{ i \vec{\alpha}(t)\cdot \vec{s}})\langle\phi_{\vec{s}}|
\end{eqnarray}

We can write the Coulomb-Volkov Hamiltonian $H_{CV}(t)$ with the help of  the operator $\vec{\pi}_{c}$,
\begin{eqnarray}\label{CVHam}
H_{CV}(t) &=& \frac{\vec{p}_{op}^2}{2m} - \frac{Z e^2}{r}  + \frac{e^2 A^2(t)}{2mc^2} - \frac{e}{mc}\vec{A}(t)\cdot\vec{\pi}_{c}
\end{eqnarray}
The corresponding Schroedinger equation is
\begin{eqnarray}\label{CVEqu}
i\hbar\frac{\partial}{\partial t} \Phi_{j}(t)&=& 
(\frac{\vec{p}_{op}^2}{2m} - \frac{Z e^2}{r} + \frac{e^2 A^2(t)}{2mc^2} - \frac{e}{mc}\vec{A}(t)\cdot\vec{\pi}_{c})\Phi_{j}(t)\nonumber\\
\end{eqnarray} 
The complete set of linearly independent solutions of Eq. (\ref{CVEqu}) is
\begin{eqnarray}\label{CVSol}
|\Phi_{j}(t) \rangle &=& e^{-\frac{i}{\hbar}\int^t (E_j + \frac{e^2A^2(t')}{2mc^2})dt' + \frac{i}{\hbar} \vec{\alpha}(t)\cdot \vec{\pi}_{c}} |\phi_j\rangle
\end{eqnarray} 
where $j \equiv \vec{p}$, stands for the momentum $\vec{p}$ of the Coulomb wave state $|\phi_{\vec{p}}^{(-)}\rangle$
and $j \equiv D$ stands for the discrete indices of the bound states $|\phi_{D}\rangle$ of the Coulomb potential.

To establish that Eq.(\ref{CVSol}) indeed satisfies Eq. (\ref{CVEqu}),
let us first consider the case  $\{j \equiv \vec{p}\}$ and use Eq. (\ref{TOpe2}) to calculate, 
\begin{eqnarray}\label{Cal1}
e^{\frac{i}{\hbar} \vec{\alpha}(t)\cdot {\vec{\pi}_{c}}} |\phi_{\vec{p}}\rangle 
&=&T(\vec{\pi}_{c})|\phi_{\vec{p}}\rangle \nonumber\\
&=& |\phi_{\vec{p}}\rangle- \sum_{\vec{s}}|\phi_{\vec{s}}\rangle (1- e^{\frac{i}{\hbar}\vec{\alpha}(t)\cdot \vec{s}} ) \langle \phi_{\vec{s}}|\phi_{\vec{p}}\rangle \nonumber\\
&=& |\phi_{\vec{p}}\rangle - |\phi_{\vec{p}}\rangle (1 - e^{\frac{i}{\hbar}\vec{\alpha}(t)\cdot \vec{p} }) \nonumber\\ 
&=& e^{\frac{i}{\hbar}\vec{\alpha}(t)\cdot \vec{p}}|\phi_{\vec{p}}\rangle 
\end{eqnarray}
Also we have
\begin{eqnarray}
- \frac{e}{mc}\vec{A}(t)\cdot\vec{\pi}_{c}|\phi_{\vec{p}}\rangle &=&-\frac{e}{mc}\vec{A}(t)\cdot \vec{p}|\phi_{\vec{p}}\rangle
\end{eqnarray}
Thus, substituting Eq. (\ref{CVSol}) in Eq. (\ref{CVEqu}) for the continuum case we get on the left hand side 
\begin{eqnarray}
l.h.s. &=& e^{-\frac{i}{\hbar} (\int^t (E_p + \frac{e^2A^2(t')}{2mc^2})dt' - \vec{\alpha}(t)\cdot\vec{p})}\nonumber\\
&\times& (E_p + \frac{e^2A^2(t)}{2mc^2} - \dot{\vec{\alpha}}(t)\cdot\vec{p})|\phi_{\vec{p}}\rangle
\end{eqnarray}
and on the right hand side
\begin{eqnarray}
r.h.s. &=& e^{-\frac{i}{\hbar}(\int^t (E_p + \frac{e^2A^2(t')}{2mc^2})dt' -\vec{\alpha}(t)\cdot\vec{p})} \nonumber\\
&\times&((\frac{{\vec{p}_{op}}^2}{2m} -\frac{Ze^2}{r})  +\frac{e^2A^2(t')}{2mc^2} - \dot{\vec{\alpha}}\cdot\vec{p})|\phi_{\vec{p}}\rangle
\end{eqnarray}
Noting that $\dot{\vec{\alpha}}(t)=\frac{e}{mc}\vec{A}(t)$ and $ H_a|\phi_{\vec{p}}\rangle =E_p|\phi_{\vec{p}}\rangle$,
where, $E_p=\frac{p^2}{2m}$, one easily sees that the 
$l.h.s = r.h.s$ and hence the given solution is exactly fulfilled. 
In a similar way it is seen that
\begin{eqnarray}
T(\vec{\pi}_{c})|\phi_{D}\rangle
&=&|\phi_{D}\rangle - \sum_{\vec{s}}|\phi_{\vec{s}}\rangle(1- e^{\frac{i}{\hbar}\vec{\alpha}(t)\cdot \vec{s}})
\langle \phi_{\vec{s}}|\phi_{D}\rangle \nonumber\\
&=& |\phi_{D}\rangle + 0 
\end{eqnarray}
since, the overlap integral between the discrete and the continuum eigenstates of the Coulomb Hamiltonian vanish by orthogonality,
$\langle \phi_{\vec{s}}|\phi_{D}\rangle=0$. 
Hence, on substituting  Eq. ($\ref{CVSol}$) in Eq.($\ref{CVEqu}$) in the discrete case we get 
\begin{eqnarray}
l.h.s.&=&e^{-\frac{i}{\hbar}\int^t (E_D + \frac{e^2A^2(t')}{2mc^2} dt' + 0)} \nonumber\\
&\times&(E_D + \frac{e^2A^2(t)}{2mc^2} + 0)|\phi_{D}\rangle
\end{eqnarray}
and
\begin{eqnarray}
r.h.s.&=& e^{-\frac{i}{\hbar}(\int^t (E_D + \frac{e^2A^2(t')}{2mc^2})dt' + 0)}\nonumber\\
&\times& ((\frac{{\vec{p}_{op}}^2}{2m} -\frac{Ze^2}{r})  
+\frac{e^2A^2(t')}{2mc^2}  +0)|\phi_{D}\rangle
\end{eqnarray}
Moreover, $(\frac{\vec{p}_{op}^2}{2m} -\frac{Ze^2}{r})|\phi_{D}\rangle =E_D|\phi_{D}\rangle$ and, hence, the $l.h.s=r.h.s$ and the verification is complete. 

To summarise, the complete set of solutions of the CV-Schroedinger equation defined by (\ref{CVEqu}) is given by Eq.(\ref{CVSol}) or, more expicitly by
\begin{eqnarray}\label{CouVolSolCon}
\Phi_{\vec{p},D}^{(-)}(\vec{r},t) = \phi_{\vec{p},D}(\vec{r})e^{-\frac{i}{\hbar} \int^t (\frac{p^2}{2m} + \frac{A(t')^2}{2mc^2} - (\frac{e}{c}\vec{A}(t')\cdot {\vec{p}) \delta_{\vec{p},D}}) dt'}\nonumber\\
\end{eqnarray} 
where for the continuum states $|\phi_{\vec{p}}\rangle$ of momentum $\vec{p}$ one has the Coulomb waves (\ref{CouWav})
and, for the discrete states $|\phi_D\rangle$ one has the well known bound states of the hydrogenic atom, 
\begin{eqnarray}
\phi_{D\equiv (n l m)}(\vec{r}) &=& N_{n l} R_{n l} (r) Y_{l m} (\theta,\phi)\nonumber\\
R_{n l}(r) &=& (2\kappa_n r)^{l} e^{- \kappa_n r}  {}_1F_{1}(-n + l +1, 2 l + 2, 2 \kappa_n r) \nonumber\\
N_{n l} &=&\frac{(2 \kappa_n)^{3/2}}{\Gamma(2 l + 2)}\sqrt{\frac{\Gamma(n + l +1)}{ 2 n\Gamma(n-l)}}
\end{eqnarray}
where $\kappa_n=\frac{Z}{n a_0}=\sqrt{\frac{-2 m E_{D}}{\hbar^2}}$.

Having thus found the explicit form of both $H_{CV}(t)$, Eq. (\ref{CVHam}), and the complete set of solutions (\ref{CVSol}) (or, alternatively, Eqs. (\ref{CouVolSolCon}) 
of the Coulomb-Volkov Schroedinger equation (\ref{CVEqu}), we can explicitly express the associated Coulomb-Volkov propagator $G_{CV}(t,t')$,
\begin{eqnarray}\label{CVPro}
G_{CV}^{(\pm)} (t,t') &=& -\frac{i}{\hbar}\theta(t-t')\nonumber\\
&\times& \{\sum_{\vec{p}} |\phi_{\vec{p}}^{(\pm)}\rangle 
e^{-\frac{i}{\hbar}\int_{t'}^{t} \frac{(\vec{p}-\frac{e}{c} \vec{A}(t'')^2}{2m} dt''} \langle\phi_{\vec{p}}^{(\pm)}| \nonumber\\
&+& \sum_{n l m} |\phi_{n l m}\rangle e^{-\frac{i}{\hbar}\int_{t'}^{t} (E_{nl} +\frac{e^2A^2(t'')}{2 m c^2})dt'' }\langle\phi_{n l m}| \}
\nonumber\\
\end{eqnarray}

\section{Coulomb-Volkov S-Matrix Series}
\label{sec:4} 
We are now ready to obtain the desired S-matrix  amplitude.
With the knowledge of $H_{CV}(t)$, Eq. (\ref{CVHam}), the final-state rest-interaction turns out to be,   
\begin{eqnarray}\label{FinIntVG}
V_{CV}(t) &= & H(t) -H_{CV}(t)\nonumber\\
&=&( -\frac{e}{mc}\vec{A}(t)\cdot(\vec{p}_{op} -\vec{\pi}_c) +V_{s.r.}(\vec{r}) )
\end{eqnarray}
We substitute the following quantities into the S-matrix amplitude (\ref{SMatSer0}): the initial and the final rest-interactions, $V_i$, Eq. (\ref{IniIntVG}) and $V_{CV}$, Eq. (\ref{FinIntVG}), the expansion of the full $G(t,t')$ in terms of the Volkov propagator, 
Eq. ({\ref{VolExpG}), and the corresponding rest-interaction $V_0$, Eq.(\ref{RestIntVol}).
This immediately yields, 
\begin{eqnarray}\label{SMatSer2}
S_{fi}&=& \langle\Phi_{\vec{p}}(t)|\phi_i(t)\rangle -\frac{i}{\hbar}\int dt_1\langle\Phi_{\vec{p}}(t_1)|V_i(t_1) |\phi_i(t_1)\rangle\nonumber\\
&-&\frac{i}{\hbar}\int dt_2 dt_1\langle\Phi_{\vec{p}}(t_2)|(-\frac{e}{mc}\vec{A}(t_2)\cdot (\vec{p}_{op} -\vec{\pi}_{c}) +\nonumber\\
&+& V_{s.r.}(\vec{r}_2))G_{Vol}(\vec{r}_2,t_2;\vec{r}_1,t_1)V_i(t_1)|\phi_i(t_1)\rangle\nonumber\\
&\cdots& 
\end{eqnarray}

Thus, finally, we have arrived at the systematic S-matrix series for the strong-field ionisation amplitude, that explicitly accounts for the Coulomb final state interaction
in all orders in terms of the Coulomb-Volkov final state. We quote the first three terms fully and give a simple rule for constructing all the higher order terms
\begin{equation}\label{SMatSerVG}
S_{fi} = \sum_{n=0}^{\infty}S_{fi}^{(n)}
\end{equation}
\begin{equation}\label{ZerOrdAmp}
S_{fi}^{(0)} = \langle\Phi_{\vec{p}}(\vec{r},t)|\phi_i(\vec{r},t)\rangle
\end{equation}
\begin{eqnarray}\label{FirOrdAmp}\label{CVSer1VG}
S_{fi}^{(1)} &=&  -\frac{i}{\hbar}\int dt_1\langle\Phi_{\vec{p}}(\vec{r}_1,t_1)|\nonumber\\
&\times&(-\frac{e}{mc}\vec{A}(t_1)\cdot\vec{p}_{op} +\frac{e^2 A^2(t_1)}{2m c^2})|\phi_i(\vec{r}_1,t_1)\rangle\nonumber\\
\end{eqnarray}
\begin{eqnarray}\label{SecOrdAmp}
S_{fi}^{(2)} &=& -\frac{i}{\hbar}\int dt_2dt_1
\langle\Phi_{\vec{p}}(\vec{r}_2,t_2)|\nonumber\\
&\times&(-\frac{e}{mc}\vec{A}(t_2)\cdot(\vec{p}_{op}-\vec{\pi}_{c})+V_{s.r.}(\vec{r}_2))\nonumber\\
&\times&G_{Vol}(\vec{r}_2,t_2;\vec{r}_1,t_1)\nonumber\\
&\times&(-\frac{e}{c}\vec{A}(t_1)\cdot\vec{p}_{op} +\frac{e^2A^2(t_1)}{2mc^2})|\phi_i(\vec{r}_1,t_1)\rangle
\end{eqnarray}
\begin{eqnarray}\label{ThiOrdAmp}
S_{fi}^{(3)} &=&-\frac{i}{\hbar} \int dt_3dt_2dt_1\langle\Phi_{\vec{p}}(\vec{r}_3,t_3)|\nonumber\\
&\times&(-\frac{e}{mc}\vec{A}(t_3)\cdot(\vec{p}_{op}-\vec{\pi}_{c}) +V_{s.r.}(\vec{r}_3)) \nonumber\\
&\times& G_{Vol}(\vec{r}_3,t_3;\vec{r}_2,t_2) (-\frac{ Z e^2}{r_2}  +V_{s.r.}(\vec{r}_2)) \nonumber\\
&\times&G_{Vol}(\vec{r}_2,t_2;\vec{r}_1,t_1)\nonumber\\
&\times& (-\frac{e}{mc}\vec{A}(t_1)\cdot\vec{p}_{op} +\frac{e^2A^2(t_1)}{2mc^2})|\phi_i(\vec{r}_1,t_1)\rangle
\end{eqnarray}
$\cdots$\\
where the angle brackets stand for the integration with respect to the space coordinates and
$``\cdots"$ stands for the higher orders terms that can be written down easily. The simple rule is to follow the structure of the third order term and to supply an extra intermediate factor $G_{Vol}V_0$ and an extra time integration, in each successive order, to all orders. 

\section{Strong-Field S-Matrix for Short-range Potentials}
{\it En passant} it is interesting to consider the S-matrix expansion of the strong-field amplitude for a system with an asymptotically short range potential. This can be obtained simply by taking the limit $Z=0$ in the result derived above. In this limit, the Coulomb waves $\phi_{\vec{p}}(\vec{r})$ reduce to the plane waves $e^{\frac{i}{\hbar}\vec{p}\cdot\vec{r}}$ and the Coulomb-Volkov state $\Phi_{\vec{p}}(\vec{r},t)$ (Eq. (\ref{CouVolSol})) reduces to the Volkov state (Eq. (\ref{VolSol})). Moreover, the final-state interaction in all terms, beginning with the second order term, reduces to the short range potential $V_{s.r.}(\vec{r})$ only due to the following simplification 
\begin{eqnarray}
&&\langle\vec{p}| (-\frac{e}{mc}\vec{A}(t)\cdot(\vec{p}_{op} -\vec{\pi}_{c})+V_{s.r.}(\vec{r})) \nonumber\\
&=&(-\frac{e}{mc}\vec{A}(t)\cdot (\vec{p}- \vec{p})\langle\vec{p}|+\langle{\vec{p}}|V_{s.r.}(\vec{r}))\nonumber\\
&=& \langle\vec{p}|V_{s.r.}(\vec{r})
\end{eqnarray}
Also, for $Z=0$, the intermediate  rest-interaction, $V_0$, in all terms (beginning with the second order term) reduces to the short-range potential $V_{s.r.}(\vec{r})$. 
Hence, in general, the Coulomb-Volkov S-matrix series, Eq. (\ref{SMatSer2}), goes over to the simpler series for $Z\equiv 0$: 
\begin{eqnarray}\label{ShoRanSer}
S_{fi}
&=& \langle\psi_{\vec{p}}(t)|\phi_i(t)\rangle  -\frac{i}{\hbar}  \int dt_1\langle\psi_{\vec{p}}(t_1)|V_i(t) |\phi_i(t)\rangle\nonumber\\
&-&\frac{i}{\hbar} \int dt_2 dt_1\langle\psi_{\vec{p}}(t_2)|V_{s.r.}(\vec{r}_2)G_{Vol}(\vec{r}_2,t_2;\vec{r}_1,t_1)\nonumber\\
&\times&V_i(t_1)|\phi_i(t)\rangle\nonumber\\
&-&\frac{i}{\hbar}\int dt_3dt_2 dt_1\langle\psi_{\vec{p}}(t_3)|V_{s.r.}(\vec{r}_3)G_{Vol}(\vec{r}_3,t_3;\vec{r}_2,t_2)|\nonumber\\
&\times&V_{s.r.}(\vec{r}_2)G_{Vol}(\vec{r}_2,t_2;\vec{r}_1,t_1)V_i(t_1)|\phi_i(t_1)\rangle\nonumber\\
&+& \cdots.
\end{eqnarray}
This series provides a self-consistent strong-field S-matrix expansion for the case of asymptotically neutral systems with effective core charge $Z=0$, e.g. for electron-detachment from negative ions.

\section{Coulomb-Volkov S-Matrix Series in Length Gauge}
There is no difficulty in obtaining the Coulomb-Volkov series in the so-called ``length gauge".
It can be obtained by starting with the Schroedinger equation of the interacting system in length gauge and following an analogous procedure as used above.
Or more simply, we may obtain it by a gauge transformation of the results already derived above in the velocity gauge.
For the sake of clarity we shall denote the quantities in length gauge below by the superscript $L$.
The Schroedinger equation of the interacting atom+ laser field, in the length gauge, is  
 \begin{equation}\label{LGSchEqu}
(i\hbar \frac{\partial}{\partial t} - H^{(L)}(t)) \Psi^{(L)}(t)=0
\end{equation}
where $H^{(L)}(t)$ is the total Hamlltonian of the system given by
\begin{equation}
H^{(L)}(t)= H_a - e\vec{F}(t)\cdot\vec{r}
\end{equation}
As before $H_a$ is the initial atomic Hamiltonian (\ref{AtoHam}) and,  $\vec{F}(t) \equiv  -\frac{1}{c}\dot{\vec{A}}(t)$
is the electric field strength. Obviously, the initial interaction is
\begin{eqnarray}\label{IniIntLG}
V_i^{(L)} (t)&=&H^{(L)} - H_a\nonumber\\
&=&- e\vec{F}(t)\cdot\vec{r}
\end{eqnarray}

The full wavefunction $\Psi^{(L)}(\vec{r},t)$ in the length gauge can be obtained by the gauge transformation
\begin{eqnarray}
\Psi^{(L)}(\vec{r},t) &=& e^{-\frac{i}{\hbar} \frac{e}{c} \vec{A}(t)\cdot\vec{r}}\Psi(\vec{r},t)
\end{eqnarray}
Similarly by gauge transforming the CV-Schroedinger equation given by Eq. (\ref{CVEqu}) in the velocity gauge, we determine the corresponding CV- Schroedinger equation and the CV-Hamiltonian in the length gauge.  
Thus, the gauge transformation of Eq. (\ref{CVEqu}) is
\begin{eqnarray}
&& e^{-\frac{i}{\hbar} \frac{e}{c} \vec{A}(t)\cdot\vec{r}} [i\hbar \frac{\partial}{\partial t} - H_{CV}(t)]
e^{\frac{i}{\hbar} \frac{e}{c} \vec{A}(t)\cdot\vec{r}}\nonumber\\
&=&(i\hbar \frac{\partial}{\partial t} - H_{CV}^{(L)}(t))
\end{eqnarray}
where a short calculation with $H_{CV}$, Eq.(\ref{CVHam}), explicitly gives the desired length gauge Coulomb-Volkov Hamiltonian $H_{CV}^{(L)}(t)$
\begin{eqnarray}\label{CVHamLG}
H_{CV}^{(L)}(t) &=& H_a  - e\vec{F}(t)\cdot\vec{r}  + \frac{e\vec{A}(t)}{mc}\cdot ( \vec{p}_{op} + \frac{e\vec{A}(t)}{c} -\vec{\pi}_{c}^{(L)})\nonumber\\
\end{eqnarray}
with
\begin{eqnarray} 
\vec{\pi}_{c}^{(L)} &=& e^{-\frac{i}{\hbar} \frac{e \vec{A}(t)}{c}\cdot\vec{r}} [ \vec{\pi}_{c} ] e^{\frac{i}{\hbar} \frac{e\vec{A}(t)}{c}\cdot\vec{r}}\nonumber\\
 &=&\sum_{\vec{s}}|\phi_{\vec{s}}^{(L)}\rangle \vec{s} \langle\phi_{\vec{s}}^{(L)}| 
\end{eqnarray}
and
\begin{eqnarray}
\langle\vec{r}|\phi_{\vec{s}}^{(L)}\rangle &=& e^{-\frac{i}{\hbar} \frac{e\vec{A}(t)}{c}\cdot\vec{r}}\phi_{\vec{s}}(\vec{r})
\end{eqnarray}
Therefore, the Coulomb-Volkov Schroedinger equation in length gauge is
\begin{equation}\label{CVEquLG}
( i\hbar \frac{\partial}{\partial t} - H_{CV}^{(L)}(t)) |\Phi_{j}^{(L)}(t)\rangle=0
\end{equation}
The corresponding complete set of linearly independent solutions of the CV-Schroedinger equation in length gauge
are 
\begin{eqnarray}\label{CVSolLG}
|\Phi_{j}^{(L)}(t) \rangle &=& e^{-\frac{i}{\hbar}\int^t (E_j + \frac{e^2A^2(t')}{2mc^2})dt' + \frac{i}{\hbar} \vec{\alpha}(t)\cdot \vec{\pi}_{c}^{(L)}} |\phi_{j}^{(L)}\rangle
\end{eqnarray} 
or, equivalently,
\begin{eqnarray}\label{CVSolConLG}
\Phi_{\vec{p}}^{(L)}(\vec{r},t) &=& e^{-\frac{i}{\hbar}\int^t (E_p + \frac{e^2A^2(t')}{2mc^2})dt' + \frac{i}{\hbar} \vec{\alpha}(t)\cdot \vec{p}}\nonumber\\
&\times& e^{-\frac{i}{\hbar c}\vec{A}(t)\cdot\vec{r}} \phi_{\vec{p}}^{(\pm)}(\vec{r})
\end{eqnarray}
where $\phi_{\vec{p}}^{(\pm)}(\vec{r})$ are the out-going (+) or the in-going (-) Coulomb waves and,
\begin{equation}\label{CVSonDisLG}
\Phi_{D}^{(L)}(\vec{r},t) = e^{-\frac{i}{\hbar}\int^t (E_D + \frac{e^2A^2(t')}{2mc^2})dt' } e^{-\frac{i}{\hbar c}\vec{A}(t)\cdot\vec{r}} \phi_{D}(\vec{r})
\end{equation}
where $j \equiv D$ stands for the discrete eigenfunctions $\phi_{D}(\vec{r})$ of the hydrogenic Hamiltonian.
Thus, the Coulomb-Volkov propagator in length gauge can be written down as 
\begin{eqnarray}\label{CVProLG}
G_{CV}^{(L)} (t,t') &=& -\frac{i}{\hbar}\theta(t-t')\nonumber\\
&\times& \{\sum_{\vec{p}} |\phi_{\vec{p}}^{(\pm)}\rangle g(t)
e^{-\frac{i}{\hbar}\int_{t'}^{t} \frac{(\vec{p}-\frac{e}{c} \vec{A}(t'')^2}{2m} dt''} g(t') \langle\phi_{\vec{p}}^{(\pm)}| \nonumber\\
&+& \sum_{n l m} |\phi_{n l m}\rangle e^{-\frac{i}{\hbar}\int_{t'}^{t} (E_{nl} +\frac{e^2A^2(t'')}{2 m c^2})dt'' }\langle\phi_{n l m}| \}.
\end{eqnarray}
where
$g(t)\equiv e^{-\frac{e}{c}\vec{A}(t)\cdot\vec{r}}$.

In the length gauge, the intermediate Volkov 
Hamiltonian is
\begin{equation}\label{VolHamLG}
H_{Vol}^{(L)}(t)=(\frac{\vec{p}_{op}^{2}}{2m} -e\vec{F}(t)\cdot\vec{r})
\end{equation}
and the well-known associated Volkov propagator is
\begin{eqnarray}\label{VolProLG}
G_{Vol}^{(L)}(\vec{r},t;\vec{r'},t') &=&-\frac{i}{\hbar}\theta(t-t') \nonumber\\
&\times& \sum_{\vec{p}} \frac{1}{L^{3} }e^{\frac{i}{\hbar} \vec{p}_{t}\cdot\vec{r}}e^{-\frac{i}{\hbar}\int^{t}_{t'}\frac{p_{t"}^{2}}{2m} dt''} 
 e^{-\frac{i}{\hbar} \vec{p}_{t'}\cdot\vec{r'}}\nonumber\\
 \end{eqnarray}
 where
 $\vec{p}_t \equiv \vec{p}-\frac{e}{c}\vec{A}(t)$.
 
The intermediate rest-interaction in length gauge is then
 \begin{eqnarray}\label{IntIntLG}
 V_{0}^{(L)}(t)&=& H^{(L)}(t) - H_{Vol}^{(L)}(t)\nonumber\\
  &=& (- \frac{Z e^2}{r} + V_{s.r.}(\vec{r}) )
 \end{eqnarray}
 
Finally, we may gauge transform the CV interaction Hamiltonian $V_{CV}(t)$, Eq. (\ref{FinIntVG}), given in the velocity gauge above, to get the Coulomb-Volkov interaction Hamiltonian in the length gauge
\begin{eqnarray}\label{IntFinLG}
V_{CV}^{(L)}(t)&=&
e^{-\frac{i}{\hbar} \frac{e\vec{A}(t)}{c}\cdot\vec{r}} [ V_{CV}(t)]
e^{\frac{i}{\hbar} \frac{e\vec{A}(t)}{c}\cdot\vec{r}}\nonumber\\
&=&-\frac{e\vec{A}(t)}{mc}\cdot (\vec{p}_{op}+\frac{e\vec{A}(t)}{c} -\vec{\pi}_{c}^{(L)})
\end{eqnarray}
where 
\begin{eqnarray}
\vec{\pi}_{c}^{(L)} = e^{-\frac{i}{\hbar c}\vec{A}(t)\cdot\vec{r}} [\vec{\pi}_{c}] e^{\frac{i}{\hbar c}\vec{A}(t)\cdot\vec{r}}
\end{eqnarray}
Having thus derived the Coulomb-Volkov Hamiltonian, Eq.(\ref{CVHamLG}), in the length gauge as well as the initial- and the final-state rest interaction Hamiltonians, Eqs. (\ref{IniIntLG}) and (\ref{IntFinLG}), and using the Volkov propgator, $G_{Vol}^{(L)}$, and associated intermediate interaction, $V_0^{(L)}$, we can now obtain the Coulomb-Volkov S-matrix series in the length gauge in an exactly analogous way as before. We may therefore simply quote the final result below,
\begin{equation}\label{SMatSerLG}
S_{fi}^{(L)} = \sum_{n=0}^{\infty}S_{fi}^{(L;n)}
\end{equation}
\begin{equation}\label{ZerOrdAmpLG}
S_{fi}^{(L;0)} = \langle\Phi_{\vec{p}}^{(L)}(\vec{r},t)|\phi_i(\vec{r},t)\rangle
\end{equation}
\begin{equation}\label{FirOrdAmpLG}\label{CVSer1LG}
S_{fi}^{(L;1)} =  -\frac{i}{\hbar}\int dt_1\langle\Phi_{\vec{p}}^{(L)}(\vec{r}_1,t_1)(- e\vec{F}(t_1)\cdot\vec{r_1})|\phi_i(\vec{r}_1,t_1)\rangle
\end{equation}
\begin{eqnarray}\label{SecOrdAmpLG} 
S_{fi}^{(L;2)} &=& -\frac{i}{\hbar}\int dt_2dt_1
\langle\Phi_{\vec{p}}^{(L)}(\vec{r}_2,t_2)|\nonumber\\
&\times&(-\frac{e}{mc}\vec{A}(t_2)\cdot(\vec{p}_{op} + \frac{e\vec{A}(t_2)}{c}-\vec{\pi}_{c}^{L)})+V_{s.r.}(\vec{r}_2))\nonumber\\
&\times &G_{Vol}^{(L)}(\vec{r}_2,t_2;\vec{r}_1,t_1)(- e\vec{F}(t_1)\cdot\vec{r_1})|\phi_i(\vec{r}_1,t_1)\rangle
\end{eqnarray}
\begin{eqnarray}\label{ThiOrdAmpLG}
S_{fi}^{(L;3)} &=&-\frac{i}{\hbar} \int dt_3dt_2dt_1\langle\Phi_{\vec{p}}^{(L)}(\vec{r}_3,t_3)| \nonumber\\
&\times&(-\frac{e}{mc}\vec{A}(t_3)\cdot(\vec{p}_{op}+\frac{e\vec{A}(t_3)}{c}-\vec{\pi}_{c}^{(L)}) +V_{s.r.}(\vec{r}_3))\nonumber\\
&\times& G_{Vol}^{(L)}(\vec{r}_3,t_3;\vec{r}_2,t_2) (-\frac{ Z e^2}{r_2}  +V_{s.r.}(\vec{r}_2))\nonumber\\
&\times& G_{Vol}^{(L)}(\vec{r}_2,t_2;\vec{r}_1,t_1)(-e\vec{F}(t_1)\cdot\vec{r_1} )|\phi_i(\vec{r}_1,t_1)\rangle\\
&\cdots& \nonumber
\end{eqnarray}
As before, the angle brackets stand for the integration with respect to the space coordinates and
$``\cdots"$ stands for the higher orders terms. They follow the same pattern as the third order term but are simply extended by an extra intermediate factor $G_{Vol}^{(L)}V_{0}^{(L)}$ and an extra time integration in each successive order, to all orders. 

\section{Concluding Remarks}
We may conclude with a few remarks. \\

(a) The first order term in the present series (Eq. (\ref{CVSer1LG}) or Eq. (\ref{CVSer1VG})) reproduces the heuristic expression introduced a long time ago \cite{jai,fer} and justifies it as a lowest order contribution. \\

(b) Beginning with the second order term the present theory opens up the possibility of systematic investigations of the role of {\it final-state} Coulomb interaction in re-scattering processes in a wide range of strong-field phenomena, including the investigations of 
the low and very low energy structures \cite{blaga,qua,wu} as well as a structure observed at/near the threshold, the so-called zero-energy structure (ZES) \cite{mos}. Despite some recent theoretical progress in their interpretation (cf. e.g. \cite{faib, ros, wbe} \& related references cited therein) it appears that, these structures remains to be fully understood. Thus, for example, the specific role played by the asymptotically long-range (Coulomb) and the short-range parts of the atomic potential in their formation, their actual numbers, or the actual  ``threshold law'' of strong-field ionisation process, are yet to be well understood. \\

(e) A gauge independent form of the Coulomb-Volkov S-matrix series can be also derived using the present method. This would be done and discussed in details elsewhere. Here we simply quote the final result for the ionisation amplitude $A_{fi}$:\\
\begin{equation}\label{CVSerGI}
A_{fi} = \sum_{n=0}^{\infty}A_{fi}^{(n)} \times e^{-\frac{i}{\hbar} E_i t_i}
\end{equation}
\begin{equation}\label{ZerOrdAmp}
A_{fi}^{(0)} = \langle\Phi_{\vec{p}}(\vec{r},t_i)\Phi_i(\vec{r},t_i)\rangle
\end{equation}
\begin{eqnarray}\label{FirOrdAmp}\label{CVSer1GI}
A_{fi}^{(1)} &=&  -\frac{i}{\hbar}\int dt_1\langle\Phi_{\vec{p}}(\vec{r}_1,t_1)|\nonumber\\
&\times&( -\frac{e}{mc}\vec{A}(t_1)\cdot(\vec{p}_{op} - \vec{\pi}_c) + V_{s.r.}(\vec{r}_1))|\Phi_i(\vec{r}_1,t_1)\rangle\nonumber\\
\end{eqnarray}
\begin{eqnarray}\label{SecOrdAmp}
A_{fi}^{(2)} &=& -\frac{i}{\hbar}\int dt_2dt_1
\langle\Phi_{\vec{p}}(\vec{r}_2,t_2)|\nonumber\\
&\times&(-\frac{e}{mc}\vec{A}(t_2)\cdot(\vec{p}_{op}-\vec{\pi}_{c})+V_{s.r.}(\vec{r}_2))\nonumber\\
&\times &G_{CV}(\vec{r}_2,t_2;\vec{r}_1,t_1)\nonumber\\
&\times&( -\frac{e}{mc}\vec{A}(t_1)\cdot(\vec{p}_{op} - \vec{\pi}_c) + V_{s.r.}(\vec{r}_1))\nonumber\\
&\times&|\Phi_i(\vec{r}_1,t_1)\rangle
\end{eqnarray}
\begin{eqnarray}\label{ThiOrdAmp}
A_{fi}^{(3)} &=&-\frac{i}{\hbar} \int dt_3dt_2dt_1\langle\Phi_{\vec{p}}(\vec{r}_3,t_3)|\nonumber\\
&\times&(-\frac{e}{mc}\vec{A}(t_3)\cdot(\vec{p}_{op}-\vec{\pi}_{c}) +V_{s.r.}(\vec{r}_3)) \nonumber\\
&\times & G_{CV}(\vec{r}_3,t_3;\vec{r}_2,t_2) \nonumber\\
&\times&( -\frac{e}{mc}\vec{A}(t_2)\cdot(\vec{p}_{op} - \vec{\pi}_c) + V_{s.r.}(\vec{r}_2))\nonumber\\
&\times& G_{CV}(\vec{r}_2,t_2;\vec{r}_1,t_1)\nonumber\\
&\times& ( -\frac{e}{mc}\vec{A}(t_1)\cdot(\vec{p}_{op} - \vec{\pi}_c) + V_{s.r.}(\vec{r}_1))|\Phi_i(\vec{r}_1,t_1)\rangle\nonumber\\
\end{eqnarray}
$\cdots$\\
\noindent where the angle brackets stand for the integration with respect to the space coordinates and
$``\cdots"$ stands for the higher orders terms that can be written down easily. The simple rule is to follow the structure of the third order term and to supply an extra factor ``$G_{CV}(\vec{r}_{n+1},t_{n+1};\vec{r}_{n},t_{n})V_f(\vec{r}_n,t_n)$", along with the corresponding space-time integration (over $d\vec{r}_{n} dt_{n+1}$), for each successive $n = 3, 4, 5,....$ to all orders, where, we note that
\begin{eqnarray}
G_{CV} (\vec{r},t;\vec{r}',t')&=& -\frac{i}{\hbar}\theta(t-t')\nonumber\\
&\times&\sum_{j,\vec{k}} \Phi_{j,\vec{k}}(\vec{r},t) \Phi_{j,\vec{k}}^*({\vec{r}}',t')|\\
\Phi_{j ,\vec{k}}(\vec{r},t)\rangle &=& e^{-\frac{i}{\hbar} \int^t (E_{j,\vec{k}} + \frac{e^2A^2(t')}{2mc^2} -(\frac{e}{mc}\vec{A}(t')\cdot \vec{k}) \delta_{j,\vec{k}} dt'}\nonumber\\
&\times& \phi_{j,\vec{k}}(\vec{r})\\
V_f (\vec{r},t) &= & -\frac{e}{mc}\vec{A}(t)\cdot(\vec{p}_{op} - \vec{\pi}_c)+ V_{s.r.}(\vec{r})\\
|\Phi_{i}(t)\rangle &=& e^{-\frac{i}{\hbar} \int^{t} (E_i + \frac{e^2A^2(t')}{2mc^2}) dt'} |\phi_i\rangle\\
|\phi_{i}(t)\rangle &=& e^{-\frac{i}{\hbar} E_i t} |\phi_i\rangle
\end{eqnarray}
and
$\vec{\pi}_c$ is given by Eq. (\ref{PIOpe}).
Above, we have set the values of the vector potential $\vec{A}(t_i) = \vec{A}(t_f)=0$ (and $\vec{F}(t_{i})=\vec{F}(t_{f})=0$) at the beginning and at the end of the laser pulse.\\

(d) The explicit expression of the Coulomb-Volkov propagator derived here (Eq. (\ref{CVPro}) or ((\ref{CVProLG})) as well as the series
(\ref{CVSerGI}) suggest that the theory could be used, as needed, to clarify the role of strong-field excitation processes involving the {\it discrete} states, either as the final state or as the mediating intermediate states or, both. For example, it could be used to gain insight into the problem of ``frustrated" vs. increased ionisation, observed some time ago {e.g. \cite{san}) and more recently with respect to  ionisation near the threshold (e.g. \cite{wu}).\\

(e) The probability of ionisation ($ P_{fi}(\vec{p})$) by an ultra-short pulse of duration $t_f-t_i$, is given simply by the absolute square of the amplitudes derived above:
\begin{eqnarray}
P_{fi}(\vec{p})= |S_{fi}|^{2} (or |A_{fi}|^2)
\end{eqnarray}
For a long pulse (with an effectively constant field amplitude) it is useful to first Fourier transform the periodic part of the S-matrix amplitude and rewrite
\begin{eqnarray}
S_{fi} (t) &=&-\frac{i}{\hbar} \sum_{n=-\infty}^{\infty} \int_{t_i}^{t_f} dt e^{\frac{i}{\hbar}(\frac{p^2}{2m} + Up + |E_i|- n\hbar\omega)t} T^{(n)}_{fi}(\vec{p})\nonumber\\
\end{eqnarray}
Then the quantity of interest is the ionisation rate (i.e. the probability of ionisation per unit interaction time ($t_f-t_i >> \frac{2\pi}{\omega}$) which can be determined from the formula
\begin{eqnarray}
R_{fi}(\vec{p})&=&\sum_{n=n0}^{\infty} \sum_{\vec{p}}\frac{2\pi}{\hbar} |T^{(n)}_{fi} (\vec{p})|^2 \nonumber\\
&\times& \delta(\frac{p^2}{2m}+ U_{p}+|E_i| - n\hbar \omega)
\end{eqnarray}
where $n_0= [\frac{\frac{p^2}{2m} + U_{p} + |E_i|)}{\hbar \omega}] _{int.}+1$, $U_{p}=\frac{e^2 F^2}{4m\omega^2}$, is the ponderomotive energy, $F$ is the peak field strength and $\omega$ is the laser frequency.

\end{document}